# Study on some interconnecting bilayer networks

Yan-Qin Qu, Xiu-Lian Xu, Shan Guan, Kai-Jun Li, Si-Jun Pan, Chang-Gui Gu, Yu-Mei Jiang, Da-Ren He*

*College of Physics Science and Technology, Yangzhou University, Yangzhou, 225002, P. R. China*

We present a model, in which some nodes (called interconnecting nodes) in two networks merge and play the roles in both the networks. The model analytic and simulation discussions show a monotonically increasing dependence of interconnecting node topological position difference and a monotonically decreasing dependence of the interconnecting node number on function difference of both networks. The dependence function details do not influence the qualitative relationship. This online manuscript presents the details of the model simulation and analytic discussion, as well as the empirical investigations performed in eight real world bilayer networks. The analytic and simulation results with different dependence function forms show rather good agreement with the empirical conclusions.

*PACS:* 89.75.Hc; 89.40.-a; 89.65.−s

*Key words*: bilayer Network; interdependent network; degree; betweenness

## 1. Introduction

In recent 10 years, people gradually accepted complex networks as the best description tool for complex systems. However, most of the discussed networks contain one kind of nodes with a certain interaction. We may address such networks as network layers. A complex system contains many interdependent nodes and their interactions. Therefore, we should describe a complex system by many network layers; each pair of them interacts also via some network layers [1,2]. Some scientists expressed the idea as *a network of networks* [3-5], a *supernetwork* [6], or *interdependent networks* [7,8]. For example, to fully describe the biological processes within the cells of human being, *E. coli* or yeast, we need many network layers, including metabolic network, protein interaction network, and gene regulatory network, etc.. Similarly, a transportation system includes airplane network, train network, coach network, etc. Each layer represents a description from one viewpoint or for a part of structure and function of the complex system.

Although the study on *network of networks* is important, often it is hard due to lack of data and suitable mathematic tools. We may perform study on several layers as the first step. As an example, Kurant and Thiran proposed a bilayer model. The lower layer describes the physical infrastructure. The upper layer describes the traffic flows in the physical infrastructure [1,2]. This article presents a new model called "interconnecting bilayer network" where some nodes appear in both layers with different functions. It is common in real world complex systems that one element performs two or even more functions. For example, a person may be a government president and simultaneously a musician. A city may contain an airport and a railway station (if we defined cities as the corresponding network nodes). They are common nodes or "interconnecting nodes (IN)" in the corresponding bilayer network (administration relationship network and musician network; airplane network and train network). Our central idea is that the "interconnecting node topological position difference (INTPD)" in the two layers, which is described by differences of some network topological properties, obeys a universal monotonically increasing dependence on "function difference (FD)" of both layers, while the interconnecting node number (INN) obeys a universal monotonically decreasing dependence on FD. As examples, the most important hub, the president, in the administration relationship network, probably cannot be the most famous musician who is the most important hub in musician network because the FD is very large. Differently, a hub airport tends to locate in a common city with a hub railway station because the FD is small. Obviously, it is meaningful to reveal the universal dependence functions if the idea is correct. However, since it is hard to quantify and measure the FD, it would be useful to present a model, in which the FD appears as a parameter, and then deduce, by the model, the relationship between INTPD and INN for a comparison with the empirical results.

For a complex system containing two network layers which have the node sets $V_1=\{i_1,i_2,\cdots,i_{M1}\}$ and $V_2=\{j_1,j_2,\cdots,j_{M2}\}$, respectively, if $V_1 \cap V_2 \neq \varphi$, we define the nodes in the nonempty intersection as IN, and the bilayer system as an "interconnecting bilayer network". In literature, many network statistical topological properties have been investigated [9]. Some of them are suitable for a description of INTPD. In our opinion, node degree, which is defined as the number of the neighboring edges of a node, should be the most suitable one. Betweenness, which is defined as the number of the shortest paths passing through the node, may be the second choice. As a try, we also choose averaged nearest neighbor degree (ANND) of an interconnecting node as the third one. We use $x$ to denote one of the three properties. We define INTPD described by $x$ as

$$U_x = \frac{1}{m}\sum_1^m \left| \frac{x_i^1}{\langle x \rangle^1} - \frac{x_j^2}{\langle x \rangle^2} \right|, \qquad (1)$$

where $x_i^1$ represents $x$ value of an IN, $i$, in the first layer, $x_j^2$ represents $x$ value of the same IN, but with a sequence number $j$ in the second layer, $\langle x \rangle^1$ denotes the averaged $x$ value of all the nodes in the first layer, $\langle x \rangle^2$ denotes the averaged $x$ value of all the nodes in the second layer, and $m$ is INN. The normalized INN (NINN) is

expressed as

$$n = \frac{m}{M_1 + M_2 - m},\qquad(2)$$

where $M_1$ denotes the total node number of the first layer, $M_2$ denotes the total node number of the second layer,. Obviously, $M_1+ M_2+m$ denotes the total node number of the two layers.

## 2. Empirical Investigation

We empirically investigated distributions, $P(x)$, of all the three properties, $U_x$ (including $U_k$, INTPD described by degree, $U_b$, INTPD described by betweenness, and $U_{knn}$, INTPD described by ANND) and NINN, $n$, in eight real world bilayer networks.

Among 39 (this number will be explained below) distributions, 34 can be fitted by "shifted power law (SPL)" functions, which are expressed as $P(x) \propto (x+\alpha)^{-\gamma}$ [10]. If $\alpha=0$, it becomes a power law; it approaches an exponential function when $\alpha\to 1$ and $x$ is normalized [11]. So, the function interpolates between a power law and an exponential decay. If we ignore the five exceptions (all of them are ANND distributions as will be reported below), we can denote each property distribution by two parameters, $\alpha$ and $\gamma$. In the following, the real world bilayer networks are labeled from 1 through 8 in increasing order of $n$, i.e. $n_1 \leq n_2 \leq \cdots \leq n_8$.

### 2.1 Empirical Investigation on Chinese herb prescription-Chinese cooked dishes bilayer network (HP-CD)

**Bilayer network 1:** In the first layer, we define herbs as nodes; two nodes are connected by an edge if they appear in at least one common prescription [10]. The data contain 917 prescriptions and 1616 herbs. There are 23035 edges between the nodes. The data were collected from a book [12], which collects the main ancient and present herb prescriptions. In the second layer, foods are defined as nodes, two nodes are connected if they appear in at least one common cooked dish [13]. 534 cooked dishes and 595 foods were included in the data, which were collected from another book [14]. There are 7876 edges between the nodes. The book collects the famous family applicable cooked dishes. An IN simultaneously is an herb and a food. 43 INs were found in the data.

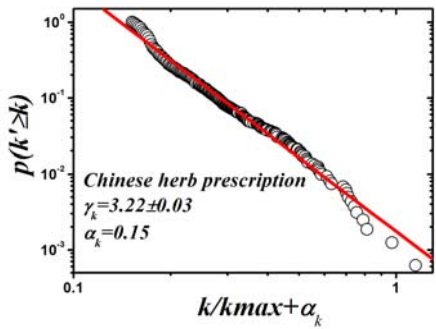 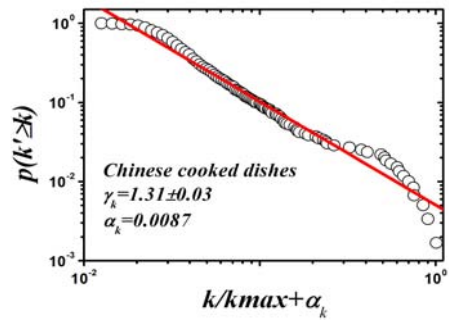

Fig. 2.1.1 Cumulative distribution of degree in herb layer.    Fig. 2.1.2 Cumulative distribution of degree in food layer.

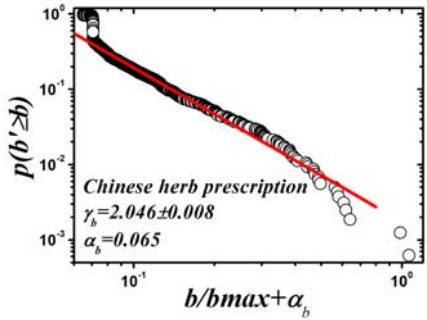    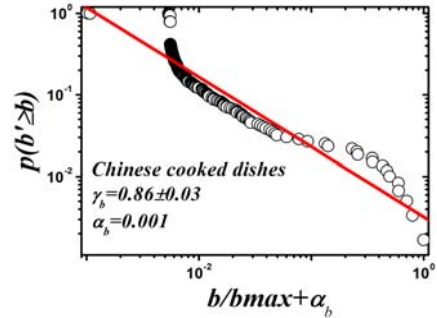

Fig. 2.1.3 Cumulative distribution of betweenness in herb layer.    Fig. 2.1.4 Cumulative distribution of betweenness in food layer.

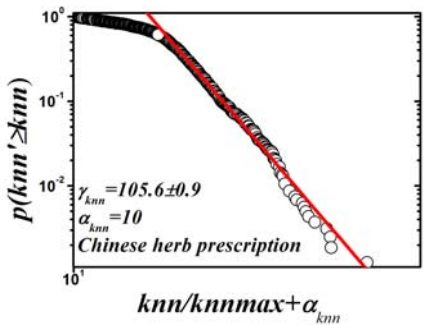    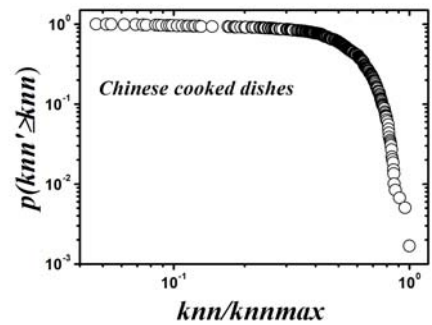

Fig. 2.1.5 Cumulative distribution of ANND in herb layer.    Fig. 2.1.6 Cumulative distribution of ANND in food layer.

We show cumulative distributions of degree, betweenness and ANND in both the layers by six figures. Degree, betweenness and averaged nearest neighbor degree are normalized by their maximum values, which are labeled in the figures by *kmax*, *bmax*, or *knnmax*, respectively, for a comparison. As can be seen, five of the distributions can be fitted by SPL functions. Only the ANND distribution in the food layer is an exception.

**2.2 Empirical Investigation on Biology keyword - Physics keyword bilayer network (BK-PK)**

**Bilayer network 2:** We define key words as nodes in both the layers. Two nodes are connected by an edge if they appear in at least one common scientific paper. The data contain 1416 biology papers and 4495 biology key words. There are 11183 edges between the nodes. The data contain 1037 physics papers and 3029 physics key words. There are 5166 edges between the nodes. An IN simultaneously is a biology key word and a physics key word. 169 INs were found in the data, which were downloaded from http://wulixb.iphy.ac.cn/cn/ch/index.aspx and http://lunwen.cnetnews.com.cn/. The data include the key words of all the papers published in a famous Chinese physics journal, "Acta Physica Sinica", in 2005 and the key words of all the Chinese biology conference papers published in 2005.

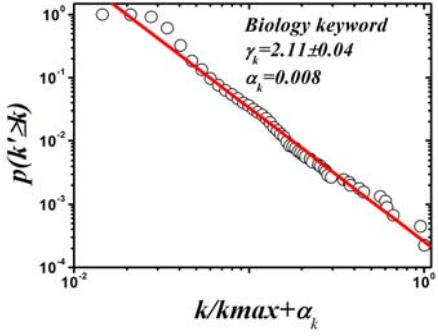
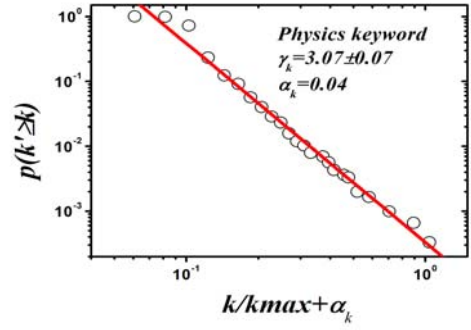

Fig. 2.2.1 Cumulative distribution of degree in biology layer.   Fig. 2.2.2 Cumulative distribution of degree in physics layer.

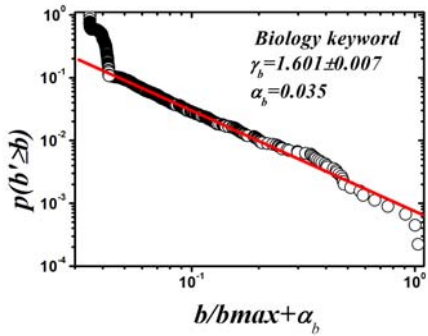
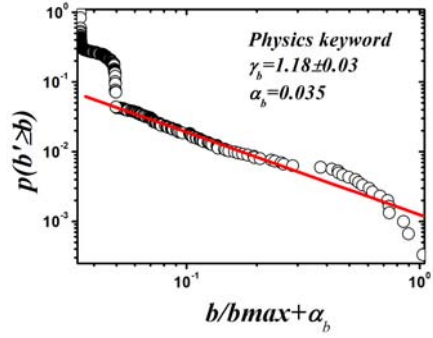

Fig. 2.2.3 Cumulative distribution of betweenness in biology layer.   Fig. 2.2.4 Cumulative distribution of betweenness in physics layer.

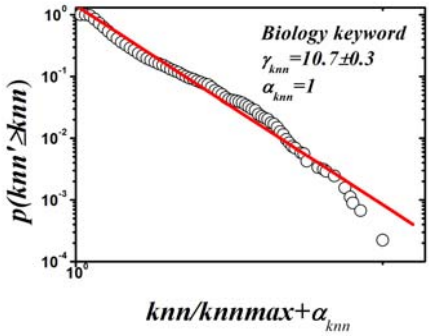
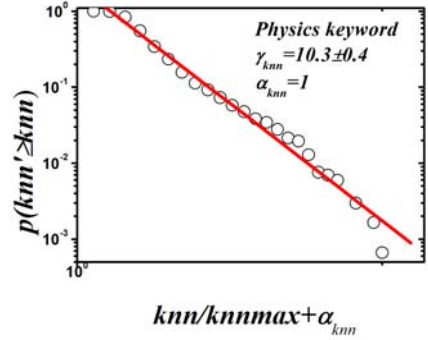

Fig. 2.2.5 Cumulative distribution of ANND in biology layer.   Fig. 2.2.6 Cumulative distribution of ANND in physics layer.

We show cumulative distributions of degree, betweenness and averaged nearest neighbor degree in both the layers by six figures. Degree, betweenness and ANND are normalized by their maximum values. All the distributions can be fitted by SPL functions.

## 2.3 Empirical Investigation on Chinese mainland movie actor - Chinese Hong Kong movie actor bilayer network (MMA-HKMA)

**Bilayer network 3:** We define movie actors as nodes in both the layers. Two nodes are connected by an edge in the first layer if they perform in at least one common mainland company movie or in the second layer if they perform in at least one common Hong Kong company movie in 2005 and 2006. The data contain 869 mainland movies and 3226 mainland movie actors. There are 44153 edges between the nodes. The data contain 337 Hong Kong movies and 1133 Hong Kong movie actors. There are 11455 edges between the nodes. An IN performs in both mainland movies and Hong Kong movies in the two years. 363 INs were found in the data, which were downloaded from http://www.cnmdb.com and http://mdbchina.com.

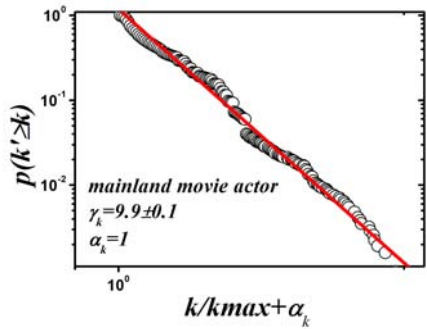 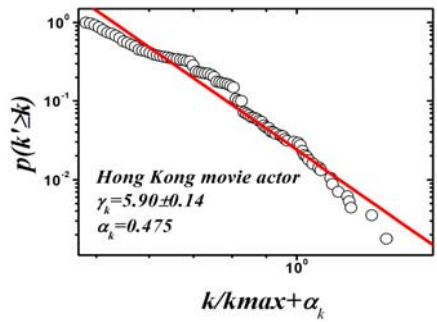

Fig. 2.3.1 Cumulative distribution of degree in mainland layer.    Fig. 2.3.2 Cumulative distribution of degree in HK layer.

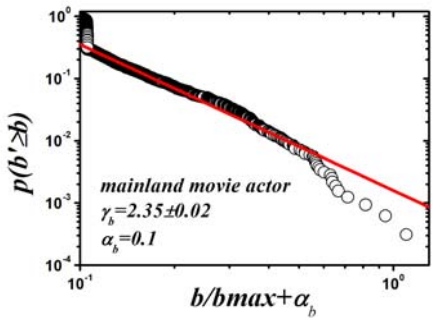 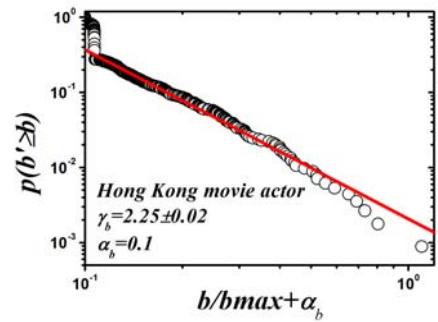

Fig. 2.3.3 Cumulative distribution of betweenness in mainland layer.    Fig. 2.3.4 Cumulative distribution of betweenness in HK layer.

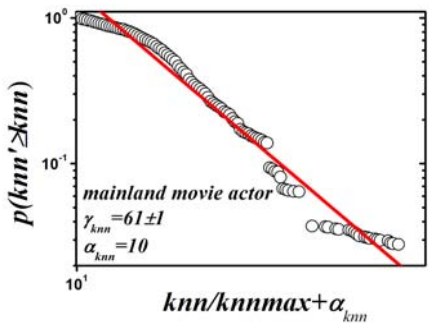 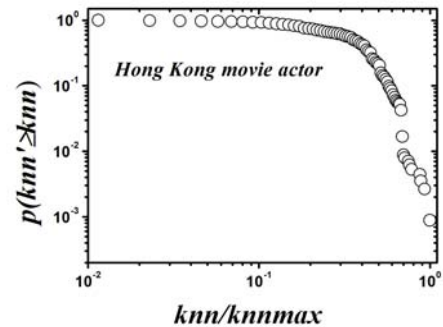

Fig. 2.3.5 Cumulative distribution of ANND in mainland layer.    Fig. 2.3.6 Cumulative distribution of ANND in HK layer.

We show cumulative distributions of degree, betweenness and averaged nearest neighbor degree in both the layers by six figures. Degree, betweenness and ANND are normalized by their maximum values. Five of the distributions can be fitted by SPL functions. Only the ANND distribution in the Hong Kong movie actor layer is an exception.

**2.4 Empirical Investigation on Baker's yeast protein interaction - metabolic bilayer network (YPI-YM)**

**Bilayer network 4:** We define proteins as nodes and physical interactions between proteins as edges in the first layer [15,16]. The data contain 3985 proteins and 30677 edges. We define enzymes (a kind of proteins) as nodes and common biochemical compounds shared by two enzymatic reactions as edges in the second layer [17]. The data contain 529 enzymes and 38285 edges. An IN is defined if both protein interaction and metabolic layers share the protein. 446 INs were found in the data, which were downloaded from http://www.ebi.ac.uk/intact/ and http://www.genome.jp/.

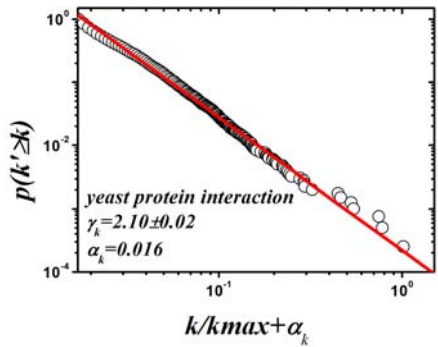
Fig. 2.4.1 Cumulative distribution of degree in YPI layer.

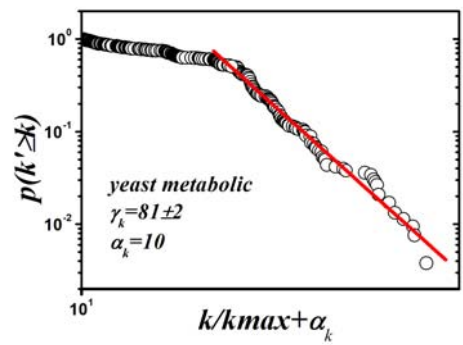
Fig. 2.4.2 Cumulative distribution of degree in YM layer.

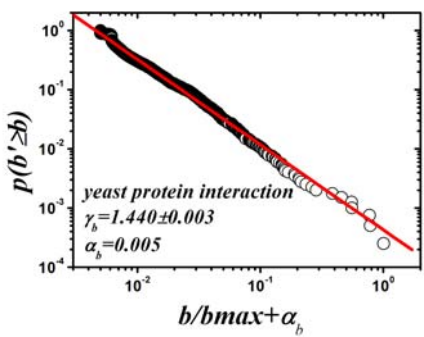
Fig. 2.4.3 Cumulative distribution of betweenness in YPI layer.

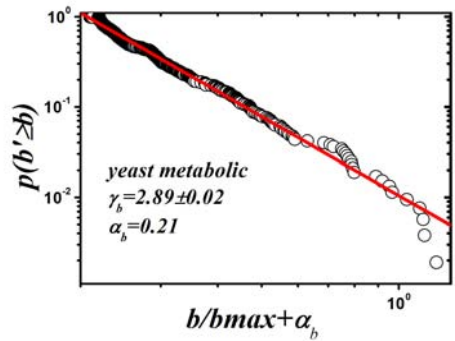
Fig. 2.4.4 Cumulative distribution of betweenness in YM layer.

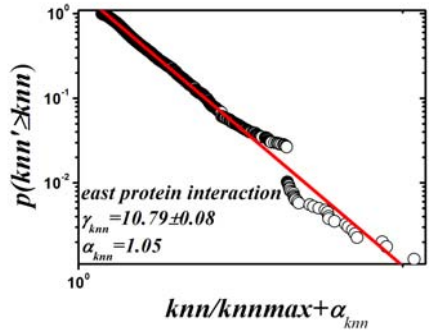 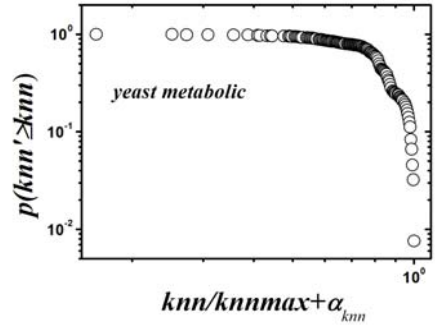

Fig. 2.4.5 Cumulative distribution of ANND in YPI layer.     Fig. 2.4.6 Cumulative distribution of ANND in YM layer.

We show cumulative distributions of degree, betweenness and averaged nearest neighbor degree in both the layers by six figures. Degree, betweenness and ANND are normalized by their maximum values. Five of the distributions can be fitted by SPL functions. Only the ANND distribution in the yeast metabolic layer is an exception.

## 2.5 Empirical Investigation on E.coli-K12 protein interaction - metabolic bilayer network (EPI-EM)

**Bilayer network 5:** The definitions and data sources are the same as in Baker's yeast protein interaction - metabolic bilayer network [15,16,17]. The data contain 2893 proteins, and 14009 edges in the first layer and 759 enzymes and **63035** edges in the second layer. 623 INs were found in the data

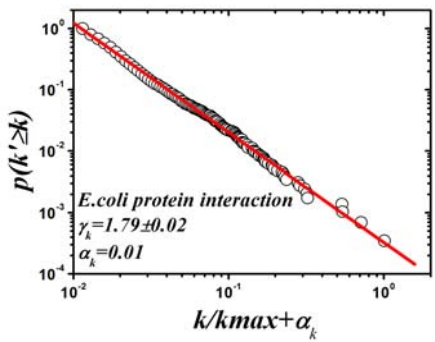 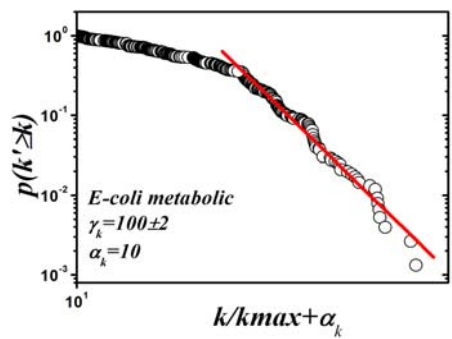

Fig. 2.5.1 Cumulative distribution of degree in EPI layer.     Fig. 2.5.2 Cumulative distribution of degree in EM layer.

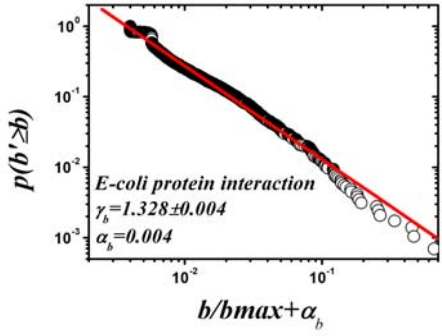 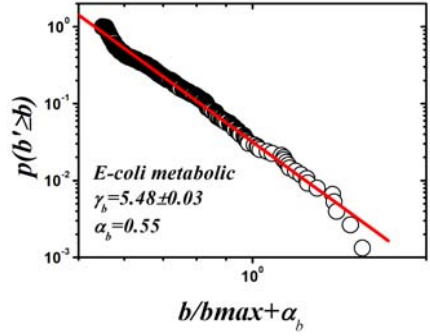

Fig. 2.5.3 Cumulative distribution of betweenness in EPI layer.    Fig. 2.5.4 Cumulative distribution of betweenness in EM layer.

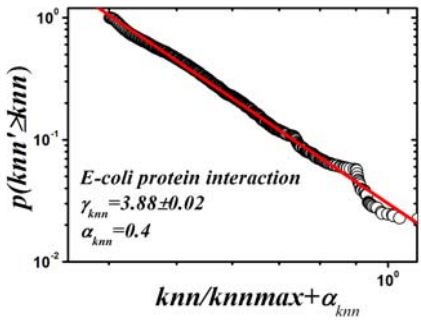 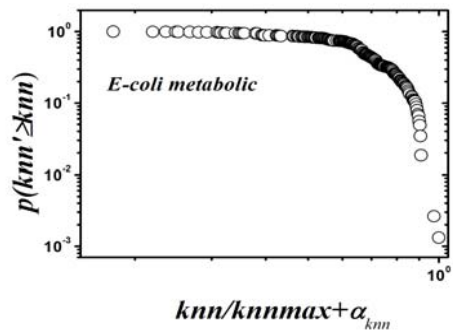

Fig. 2.5.5 Cumulative distribution of ANND in EPI layer.    Fig. 2.5.6 Cumulative distribution of ANND in EM layer.

We show cumulative distributions of degree, betweenness and averaged nearest neighbor degree in both the layers by six figures. Degree, betweenness and ANND are normalized by their maximum values. Five of the distributions can be fitted by SPL functions. Only the ANND distribution in E.coli-K12 metabolic layer is an exception.

**2.6 Empirical Investigation on Chinese Traffic bilayer networks:**

**Bilayer network 6: coach - airplane bilayer network (coach - airplane),**

**Bilayer network 7: train – airplane bilayer network (train – airplane),**

**Bilayer network 8: coach – train bilayer network (coach –train)**

We define cities containing coach stations, train stations, or airports with an administration level higher than "regional counties" (China mainland is divided into 31 provinces, which are further divided into 333 regional counties) as nodes. Two nodes are connected by an edge if corresponding traffic tool (coach, train or airplane) provides direct traffic service (without changing coach, train or airplane) between them. The data, which were downloaded from http://www.china-holiday.com, http://www.ipao.com, http://train.hepost.com/, contain 314 cities with coach stations and 3220 edges in the coach layer, 251 cities with train stations and 6775 edges in the train layer, and 100 cities with airports and 838 edges in the airplane layer. An IN is defined if a city contains two kinds of stations, i.e., a coach station and an airport, a train station and an airport, or a coach station and an train station. 100 INs were found in the coach-airplane bilayer data. 88 INs were found in the train–airplane bilayer data. 251 INs were found in the coach- train bilayer data.

Now it is clear that we investigated eight real world bilayer systems, but there are only 13 different layers. The three topological properties (degree, betweenness and averaged nearest neighbor degree) show 39 distributions.

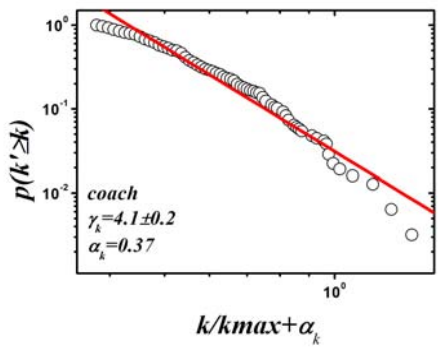
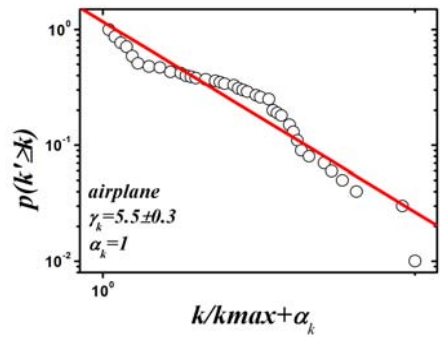

Fig. 2.6.1 Cumulative distribution of degree in coach layer.    Fig. 2.6.2 Cumulative distribution of degree in airplane layer.

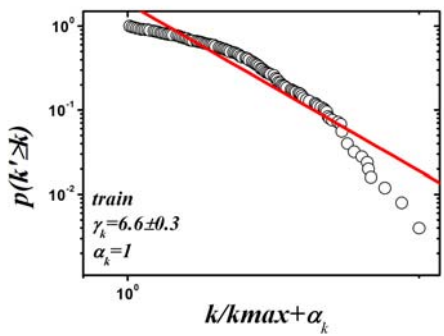
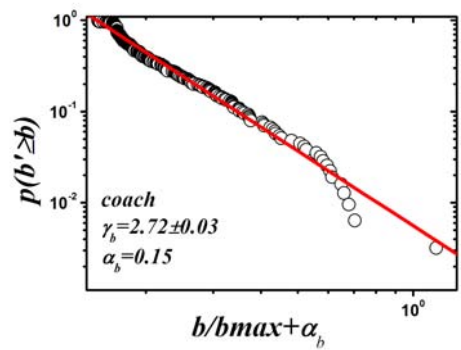

Fig. 2.6.3 Cumulative distribution of degree in train layer.    Fig. 2.6.4 Cumulative distribution of betweenness in coach layer.

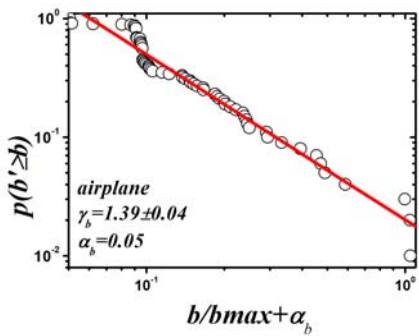
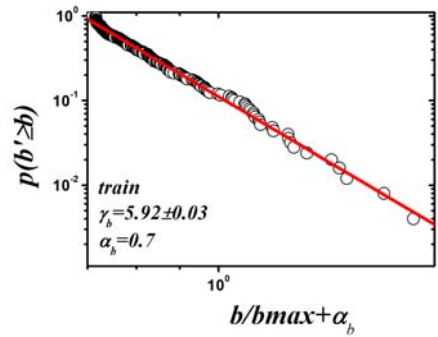

Fig. 2.6.5 Cumulative distribution of betweenness in airplane layer.    Fig. 2.6.6 Cumulative distribution of betweenness in train layer.

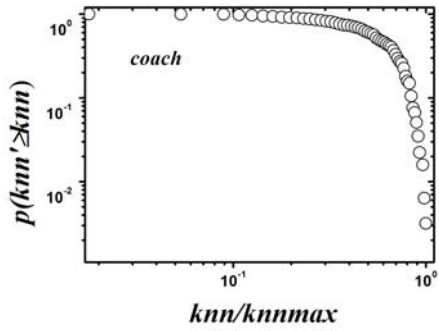 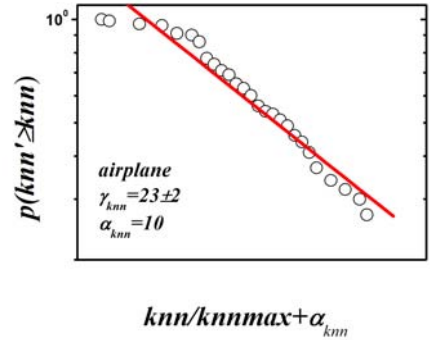

Fig. 2.6.7 Cumulative distribution of ANND in coach layer.    Fig. 2.6.8 Cumulative distribution of ANND in airplane layer.

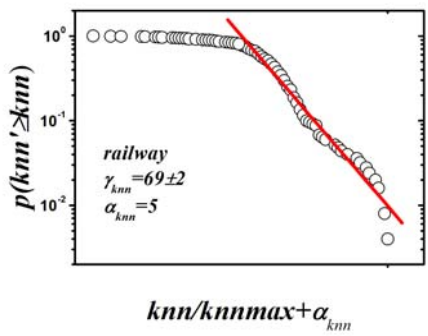

Fig. 2.6.9 Cumulative distribution of ANND in train layer.

We show cumulative distributions of degree, betweenness and averaged nearest neighbor degree in the three layers by nine figures. Degree, betweenness and ANND are normalized by their maximum values. Eight of the distributions can be fitted by SPL functions. The only exception is ANND distribution of the coach layer.

Table 1 and 2 summarize properties and parameters of all the layers and bilayer networks.

Table 1: Parameters of the layers. $M$ is the node number, $e$ denotes the edge number, $\alpha_k$, $\gamma_k$, $\alpha_b$, $\gamma_b$, $\alpha_{knn}$, $\gamma_{knn}$ denote the SPL distribution parameters of degree, betweenness, or ANND, respectively. "Not" means that the distribution cannot be fitted by SPL functions.

| Bilayer No. | Layer | $M$ | $e$ | $\alpha_k$ | $\gamma_k$ | $\alpha_b$ | $\gamma_b$ | $\alpha_{knn}$ | $\gamma_{knn}$ |
|---|---|---|---|---|---|---|---|---|---|
| 1 | HP | 1616 | 23035 | 0.15 | 3.22 | 0.065 | 2.046 | 10 | 105.6 |
| 1 | CD | 595 | 7876 | 0.0087 | 1.31 | 0.001 | 0.86 | not | not |
| 2 | BK | 4495 | 11183 | 0.008 | 2.11 | 0.035 | 1.601 | 1 | 10.7 |
| 2 | PK | 3029 | 5166 | 0.04 | 3.07 | 0.035 | 1.18 | 1 | 10.3 |
| 3 | MMA | 3226 | 44153 | 1 | 9.9 | 0.1 | 2.35 | 10 | 61 |
| 3 | HKMA | 1133 | 11455 | 0.475 | 5.9 | 0.1 | 2.25 | Not | Not |
| 4 | YPI | 3985 | 30677 | 0.016 | 2.10 | 0.005 | 1.440 | 1.05 | 10.79 |
| 4 | YM | 529 | 38285 | 10 | 81 | 0.21 | 2.89 | Not | Not |
| 5 | EPI | 2893 | 14009 | 0.01 | 1.79 | 0.004 | 1.328 | 0.4 | 3.88 |
| 5 | EM | 759 | 63035 | 10 | 100 | 0.55 | 5.48 | Not | Not |
| 6 | coach | 314 | 3220 | 0.37 | 4.1 | 0.15 | 2.72 | Not | Not |
| 6 | airplane | 100 | 838 | 1 | 5.5 | 0.05 | 1.39 | 10 | 23 |
| 7 | train | 251 | 6775 | 1 | 6.6 | 0.7 | 5.92 | 5 | 69 |
| 7 | airplane | 100 | 838 | 1 | 5.5 | 0.05 | 1.39 | 10 | 23 |
| 8 | coach | 314 | 3220 | 0.37 | 4.1 | 0.15 | 2.72 | Not | Not |
| 8 | train | 251 | 6775 | 1 | 6.6 | 0.7 | 5.92 | 5 | 69 |

Table 2: Parameters of the bilayer networks. $m$ is the IN number, $n$ denotes the ratio of IN over all the nodes, $U_k$, $U_b$, $U_{knn}$ denote the INTPD described by degree, betweenness, or ANND, respectively.

| Bilayer / Property | 1 | 2 | 3 | 4 | 5 | 6 | 7 | 8 |
|---|---|---|---|---|---|---|---|---|
| $m$ | 43 | 169 | 363 | 446 | 623 | 100 | 88 | 251 |
| $n$ | 0.0198 | 0.023 | 0.091 | 0.11 | 0.257 | 0.3185 | 0.335 | 0.8 |
| $U_k$ | 2.694 | 1.204 | 1.039 | 0.972 | 1.027 | 0.719 | 0.655 | 0.6089 |
| $U_b$ | 3.758 | 3.406 | 2.878 | 1.315 | 1.291 | 1.067 | 0.905 | 0.887 |
| $U_{knn}$ | 0.556 | 0.598 | 0.528 | 0.866 | 0.827 | 0.382 | 0.267 | 0.314 |

## 3. The model and discussions

The empirical results show that, with the increasing of $n$, the 24 $U_x$ data ($U_k$, $U_b$ and $U_{knn}$) basically show monotonic decreasing (see Table 2), which supports our central idea. However, it is impossible to extract the quantitative relationship between $n$ and $U_x$ as well as their dependences on FD just by 24 data. We have to construct a model, which quantitatively expresses the central idea, and then compare the model simulation and analytic results with the empirical data.

We can briefly interpret the model as follows. (1) Construct two network layers; in both layers $x$ parameter distributions obey SPL functions with parameters $\alpha_1$, $\gamma_1$, $\alpha_2$, $\gamma_2$, respectively. (2) Select node $i$ in the first layer with a probability $p_1=(1-\delta)^\beta$ ($0\leq\delta\leq1$) where $\delta$ represents FD and $\beta$ is an adjustable constant. This function determines NINN (how many nodes will become IN). (3) Select node $j$ in the second layer with a probability $p_2 = \delta[1/(1-u_{ij})]/\sum_i[1/(1-u_{ij})]+(1-\delta)(1/u_{ij})/\sum_i(1/u_{ij})$ where $u_{ij}=(1/u_{max})|x_i^1/\langle x\rangle^1 - x_j^2/\langle x\rangle^2|$, $u_{max}=\max\{x_{max}^1/\langle x\rangle^1, x_{max}^2/\langle x\rangle^2\}$ and $x_{max}^1$ denotes the maximum value of $x$ in the first layer. This function determines INTPD ($i$ will merge with a node in the other layer). (4) Merge the two nodes, $i$ and $j$, into one IN. We have tried different $p_2$ functions, such as $p_2 = \delta(u_{ij}^l/\sum_i u_{ij}^l)+(1-\delta)(1-u_{ij})^l/\sum_i(1-u_{ij})^l$ where $l$ is an adjustable constant or $p_2 = \delta(u_{ij}/\sum_i u_{ij})+(1-\delta)(1/u_{ij})/\sum_i(1/u_{ij})$. The simulation shows almost no difference by the three function forms. In addition, as shown in Fig. 1, the differences between the simulation results with different $\beta$ values are small. These simulation results may show that the central idea is rather tolerant of the dependence function details. This is why we can choose one of the possible function forms. (5) Repeat steps (3) and (4) until every pair of nodes, $i$ and $j$, are considered (please note that each node can merge with only one node in the other layer).

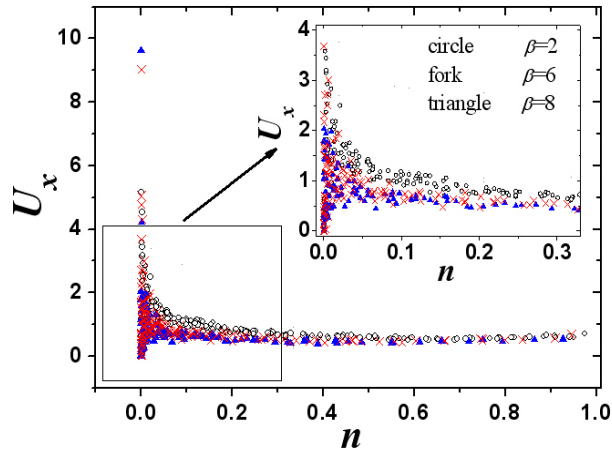

Fig.1 (color online)   Simulation results about relationship between $U_x$ and $n$ with $\beta$ values 2, 6, 8, respectively. The parameter values are taken approximately as the averaged value of the empirical data. They are: $\alpha_1=0.2$, $\gamma_1=3$, $\alpha_2=0.5$, $\gamma_2=6$, $x_{max}^1=x_{max}^2=500$, $x_{min}^1=x_{min}^2=1$, $M_1=M_2=1000$.

In order to make an analytic discussion, we have to simplify the model further. (1) Let the node number in both the layers be same, i. e., $M_1=M_2=M$, and all the three property distributions show a common function form $p(x) = c(x/x_{max} + \alpha)^{-\gamma-1}$ where $c$ is a constant for normalization and can be expressed as $c = \gamma\{x_{max}[(x_{min}/x_{max} + \alpha)^{-\gamma} - (1+\alpha)^{-\gamma}]\}^{-1}$. Therefore we have $\langle x \rangle^1 = \langle x \rangle^2 = \langle x \rangle$. (2) Let node $i$ and $j$ labeled from 1 through $M$ in increasing order of $x$, i.e. $x_1 \leq x_2 \leq \cdots \leq x_M$, set up a simplified deterministic rule for selecting IN, which is to merge $i$ and $j$ if $x_i^1 = x_j^2 + \Delta$ ($0 \leq \Delta \leq x_{max}$) where $\Delta$ is the main adjustable constant parameter representing FD. With the simplifications, $U_x$ dependence on $\Delta$ can be simplified as $U_x=\Delta/\langle x \rangle$, i.e., all the three INTPD depend on $\Delta$ in a very simplified function form. NINN, $n$, should show a monotonic decreasing dependence on $\Delta$. When $\Delta=0$, all the nodes in both the layers merge, $n=M$; while $n=1$ (only one pair of nodes merge) if $\Delta=x_{max}-x_{min}$. The dependence functions become very simple but still agree with our central idea presented at the beginning of the text. Then we can calculate NINN as a function of INTPD by $m = \sum_{x_{min}+\Delta}^{x_{max}} MP(x)$ and the definition (2). We have

$$m = \sum_{x_{min}+\Delta}^{x_{max}} MP(x) \approx \int_{(x_{min}+\Delta)}^{x_{max}} Mc(x/x_{max} + \alpha)^{-\gamma-1} dx = \gamma^{-1} x_{max} Mc[(x_{max}^{-1}(x_{min}+\Delta) + \alpha)^{-\gamma} - (1+\alpha)^{-\gamma}]$$
$$= M[(x_{max}^{-1}(x_{min}+\Delta) + \alpha)^{-\gamma} - (1+\alpha)^{-\gamma}]/[(x_{min}/x_{max} + \alpha)^{-\gamma} - (1+\alpha)^{-\gamma}] \quad (3)$$

Input (3) into definition (2), one gets NINN as

$$n = \frac{2}{2 - [((x_{min}+\Delta)/x_{max} + \alpha)^{-\gamma} - (1+\alpha)^{-\gamma}]/[(x_{min}/x_{max} + \alpha)^{-\gamma} - (1+\alpha)^{-\gamma}]} - 1 . \quad (4)$$

The average of $x$ can be calculated as

$$\langle x \rangle = \int_{x_{min}}^{x_{max}} xP(x)dx = \{\gamma x_{max}[(x_{min}/x_{max} + \alpha)^{-\gamma+1} - (1+\alpha)^{-\gamma+1}]\}/\{(\gamma-1)[(x_{min}/x_{max} + \alpha)^{-\gamma} - (1+\alpha)^{-\gamma}]\} - \alpha x_{max} . \quad (5)$$

Input (5) in the eq. $U_x=\Delta/\langle x \rangle$, one obtains

$$\Delta = U_x\{\{\gamma x_{max}[(\frac{x_{min}}{x_{max}} + \alpha)^{-\gamma+1} - (1+\alpha)^{-\gamma+1}]\}/\{(\gamma-1)[(\frac{x_{min}}{x_{max}} + \alpha)^{-\gamma} - (1+\alpha)^{-\gamma}]\} - \alpha x_{max}\} . \quad (6)$$

Input (6) into (4), we obtain

$$n = \frac{2}{2 - B_1^{-1}\{[A + U_x\gamma B_2/((\gamma-1)B_1) - U_x\alpha + \alpha]^{-\gamma} - (1+\alpha)^{-\gamma}\}} - 1 , \quad (7)$$

Where $A=x_{min}/x_{max}$, $B_1=(A+\alpha)^{-\gamma}-(1+\alpha)^{-\gamma}$, $B_2=(A+\alpha)^{-\gamma+1}-(1+\alpha)^{-\gamma+1}$.

In Fig. 2, the two wide solid lines (green online) are drawn by (7). The upper line is drawn with the parameter values $\alpha=1$, $\gamma=10.7$, $x_{min}=1$, $x_{max}=500$, which give rise to the highest position of the lines within the empirical parameter range. The lower line is drawn with the parameter values $\alpha=0$, $\gamma=5$, $x_{min}=1$, $x_{max}=50$, which give rise to the lowest position of the lines within the empirical parameter range. In Fig. 2 the much smaller cross and triangle (blue online) denote the simulation results. The triangle data were computed with the parameter values $\alpha_1 = \alpha_2 = 0.01$, $\gamma_1=\gamma_2=1.1$, $x_{min1}= x_{min2}=1$, $x_{max1}= x_{max2}= 50$, $M_1=M_2=1000$, $\beta=6$, which give rise to the lowest position of the simulation results within the empirical parameter range. The cross data were computed with the parameter values $\alpha_1 =1$, $\gamma_1=10.7$, $x_{min1}=1$, $x_{max1}=500$, $\alpha_2 =0.0087$, $\gamma_2=1.31$, $x_{min2}=1$, $x_{max2}=200$, $M_1=M_2=1000$, $\beta=6$, which give rise to the highest position within the empirical parameter range. The much larger fork, cross and triangle (red online) denote the empirical data, from system 1 to 8 with the increasing order of $n$. The fork, cross and triangle data denote $U_k$, $U_{knn}$ and $U_b$, respectively. One can see that, with such simple dependence functions, the analytic conclusion still show rather good agreement with the simulation and empirical results. Since the eight real world bilayer networks have been selected "randomly", the conclusion shown in Fig. 2 indicates the universality and tolerance of function details.

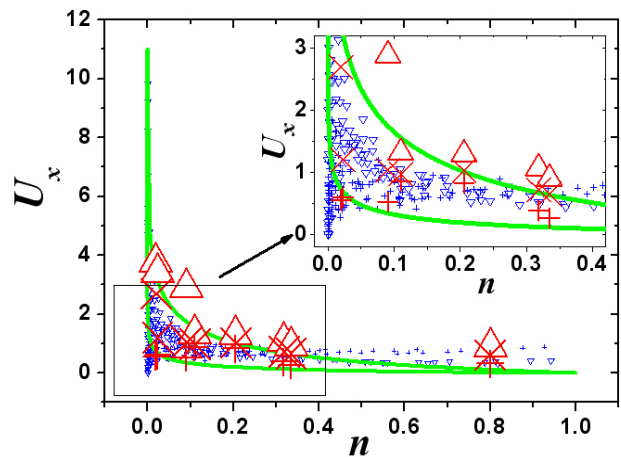

Fig.2. (color online)   Simulation, analytic and empirical results about relationship between $U_x$ and $n$.

In conclusion, we propose a model describing a new general kind of interactions (or interdependence) of two "network layers" in which merging of some nodes in the layers happens. The merge represents that the (interconnecting) nodes play the roles in both the layers. The central idea of the model is that the interconnecting node topological position difference monotonically increases and the interconnecting node number monotonically decreases when function difference of the two layers becomes increasingly larger. Our analytic and simulation results with different dependence function forms show rather good agreement with the empirical conclusions obtained in eight real world bilayer networks.

It is common that real world networks are interdependent on each other [7,8]. Some of the general interdependences have been theoretically and empirically investigated [1,2,7,8], however, there should be much more discoveries waiting for our further study. The research on this direction is very useful for understanding complex systems.

This work is supported by the Chinese National Natural Science Foundation under grant numbers 10635040